# Multi Object Spectrograph of the Fireball-II Balloon Experiment

G. R. Lemaitre, R. Grange, S. Quiret, B. Milliard, S. Pascal, V. Lamandé

*Laboratoire d'Astrophysique de Marseille, LOOM, Aix Marseille Université and CNRS,*
*38 rue Fréderic Joliot-Curie, F-13388 Marseille CX 13, France, EU*
*Email : gerard.lemaitre@lam.fr*

**Abstract:** Fireball-II is a NASA/CNES balloon-borne telescope and MOS to study faint diffuse emissions of galaxies in the space ultraviolet. The MOS is based on two identical reflective Schmidt systems sharing an plane-aspherized grating obtained by active optics methods.

## 1. Introduction

Fireball-II (Faint Intergalactic Redshifted Emission Balloon) is a NASA/CNES balloon-borne experiment to study the faint diffuse intergalactic medium from the line emissions in the ultraviolet (200nm) above 37km flight altitude. For the 2009 flight, Fireball-I relied on a fiber bundle Integral Field Unit (IFU) spectrograph fed by a 1 meter diameter parabola. This latter design used a reflective convex grating in Offner Littrow mounting [1].

As the science goals are concentrating on the circumgalactic medium, Fireball-II will use a Multi Object Spectrograph (MOS) for the 2015 flight that takes full advantage of the new high QE, low noise 13.5μm pixels UV CCD. This will increase the number of targets per flight while keeping the fast f-number ($f/2.5$) to maintain a high signal to noise ratio. The R ~ 2200 and 1.5arcsec FWHM MOS has a much larger field of view (600 arcmin$^2$) than the IFU (16 arcmin$^2$) for Fireball-I and thus we designed a two-mirror parabola field corrector to reimage the sky onto a spherical slit mask optimized to produce a flat field at the spectrograph focal plane.

After the field corrector system and beyond the mask the new spectrograph is based on two identical Schmidt systems acting as collimator and camera both sharing a 2400g/mm reflective Schmidt grating. The asperization of the grating is achieved by double replication technique of a metal deformable matrix whose quasi-constant thickness active clear aperture zone is built-in to a rigid elliptical contour. We will present the optical design of the system and the elasticity design of the deformable matrix for obtaining the aspheric grating replica.

## 2. Optical Design for Fireball-II

The light from the sky is first reflected on a 1x1.3m steerable holed plane mirror allowing field acquisition and tip-tilt slewing control during exposure. A 1m concave parabolic mirror upside down provides an $f/2.5$ focal surface – which require coma correction – through the hole of the steerable mirror.

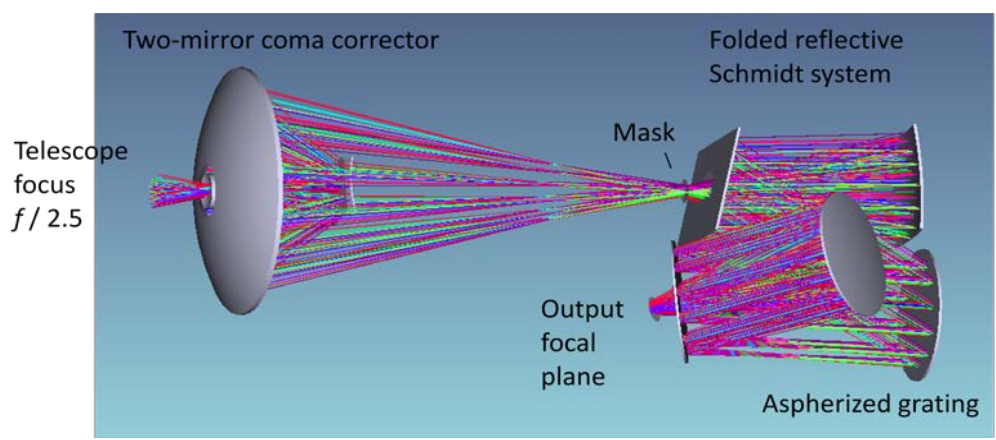

Fig. 1 Optical design of Fireball-II with two-mirror coma corrector and twin reflective Schmidt with shared aspherized grating.

The optical design of the multi-object spectrograph (MOS) is based on a *two-mirror field corrector $f/2.5$-$f/2.5$* for coma correction which provides the field of 600 arcmin$^2$. Then the output focal surface allows use of a multi-holed reflective mask for the direct input of object spots to the spectrograph and a channel from reflection on the mask for acquisition control and guiding. The *f* /2.5-*f* /2.5 spectrograph is made of two identical *folded reflective Schmidt systems* sharing a reflective aspherical grating (Fig. 1).The optical system provides a spectral resolution of 0.1nm at 200nm. Compared to Fireball-I, the new design should provide several gains in throughput : a factor 6 on QE efficiency of the detector and a factor 2 on the grating efficiency. Compared to the pixel size, the better image quality of the new design should also improve the limiting magnitude. Taking into account the increase number of optical surfaces and the gain in image quality, one expects a gain of 2.7 or 3 in limiting magnitude.

### 3. Aspherization of Reflective Gratings by Active Optics Methods

Among possible methods to achieve aberration corrections directly with a diffraction grating, our optical lab (LOOM) investigated active optics methods which are based on a double replication technique via a deformable matrix, also called *active submaster*. These methods were applied to obtain as well to elastically generate balanced a primary spherical aberration mode *Sphe*3 in combination with a curvature mode *Cv*1 – for *plane-aspheric gratings* – as to generate a primary astigmatism mode *Astm*3 – for *toroid gratings* [2].

### 3.1 Balanced primary spherical aberration $z = A_{20} r^2 + A_{40} r^4$

A combination of *Cv*1 and *Sphe*3 modes with $A_{20}A_{40} < 0$ allows reducing the stress level in the deformable substrate. Such profiles are of particular interest for any reflective systems as, for instance, high-throughput Schmidt cameras of spectrograph types equipped with aspheric reflective gratings. Balancing *Cv*1 and *Sphe*3 modes also substantially improve the optical performance by *minimizing* field residual aberrations. For reflective gratings used in normal diffraction, i.e. *β*=0, an important law, established by Lemaitre [2,3], states that:

 *A pupil reflective surface with balanced radial variation of its local curvatures minimizes the field aberrations of any reflective Schmidt system.*

For a diffraction angle *β*≠0, this law also applies to the principal directions of the elliptic pupil. For *β*=0, one shows that its best surface is of rotational symmetric and then the equation of the aspheric surface must be

$$z = z_0 (3\rho^2 - \rho^4), \qquad (1)$$

where $z_0$ is a constant and $\rho \in [0,1]$ the dimensionless aperture radius $r/r_{max}$. Hence, from (1), the optics null-power zone – no deviation of a ray – is located at $\rho = \sqrt{(3/2)} = 1.225…$ i.e. outside the clear optical aperture of diameter $2r_{max}$.

Many deformable submasters with built-in boundary at the edge, quasi-constant thickness distribution (q-CTD) and bent by a uniform loading *q* were developed at LOOM to generate aspherized gratings through the double replication technique. For fast *f*-ratios the slight decrease of the thickness from center towards edge provides the 5th-order correction *Sphe*5 (Figs. 2 and 3).

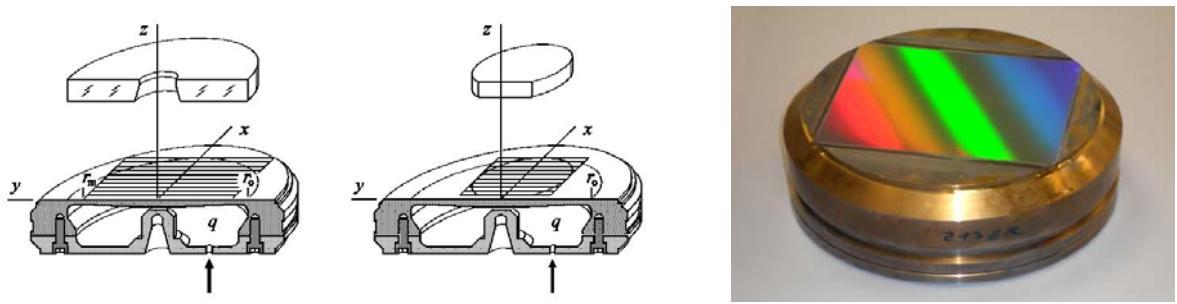

Fig. 2. *Left*: Deformable submasters of quasi-constant thickness distribution (q-CTD) for on-axis or off-axis grating aspherizations by double replicas. *Right*: View of an intermediate replica on active submaster (LOOM).

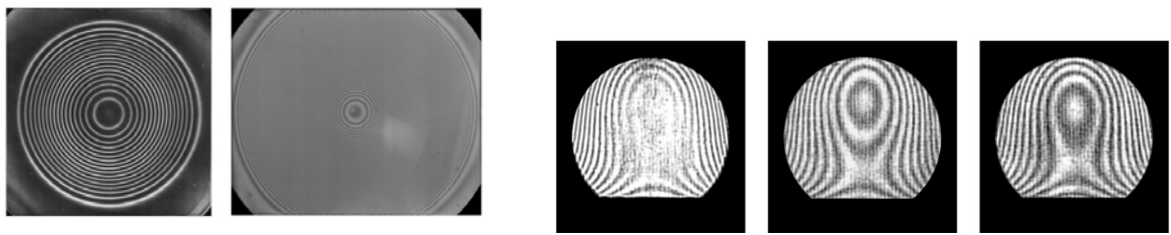

Fig. 3. He–Ne fringes with respect to flat of on- and off-axis (*freeform*) reflective grating replicas aspherized from 2nd replication on Zerodur substrate for the MARLY, CARELEC and ISARD faint object spectrographs (LOOM).

## 3.2. Primary astigmatism mode  $z = A_{22}\, r^{\,2} \cos 2\theta$

The primary astigmatism mode, *Astm*3, can be generated either from *constant thickness distribution* (CTD) and *variable thickness distribution* (VTD) mirror class [2]. The flexure to be achieved is a saddle-like surface i.e., with opposite curvatures in main orthogonal directions. Within VTD class, the basic solution is a *cycloid-like* thickness on which a net shearing force of the form $\cos 2\theta$ is applied along the mirror contour. Practicable configurations were developed at LOOM by use of either four outer bending bridges for CTDs or opposite force-pairs on simply-supported outer ring for VTDs [2]. Stainless steel submasters from the VTD class were designed with the *cycloid-like* thickness distribution $t/t_0 = (1 - \rho^2)^{1/3}$ and four 90° bridges [2]. A second stage used two opposite couple of forces for the stressing and final replication on rigid Zerodur substrates. This was developed for the *single surface* spectrographs UVCS and CDS of the SOHO Mission ESA/NASA orbiting at Lagrange point $L_1$ (Fig. 4)

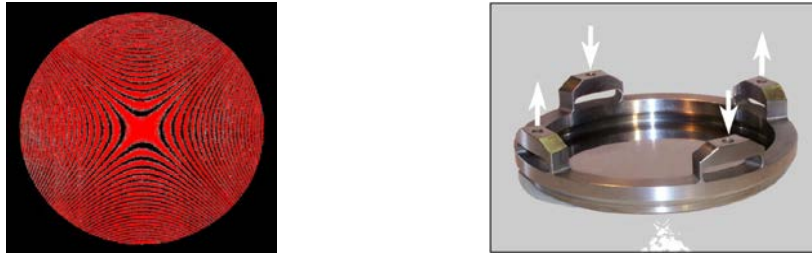

Fig. 4. *Left*: Cycloid-like deformable submaster generating *Astm*3 mode from four forces. Thickness distributions $t/t_0 = (1 - \rho^2)^{1/3}$. *Right*: He-Ne fringes of a first replica grating on active submaster for obtaining by replication one of the toroid *Astm*3-corrected grating replicas of spectrographs UVCS and CDS of SOHO Mission (LOOM, ETH-Zurich, Bach Research Corp-Boulder).

## 4. Aspherization of a Freeform Reflective Grating for Fireball-II

An active optics development to obtain the freeform surface of a reflective Schmidt first mirror was carried out by Lemaitre [4] for an $f/1.5$ prototype telescope, 20cm aperture, using a *tulip form* thickness distribution with a set of punctual axial forces. Taking into account the law stated in above section 3.1, the best optical performance is achieved when the sizes of elliptic clear aperture and elliptic null-power zone (outside) are set in the ratio $\sqrt{3/2}$. This law is equivalent in setting the *algebraic balance of local curvatures* – $\partial^2 z/\partial x^2$ and $\partial^2 z/\partial y^2$ – in the principal directions. It was retained for the design of the giant-segmented telescope LAMOST. From (1), the equation of the stressed matrix surface, of opposite sign to that of the grating surface, becomes in $3^{rd}$-order [2]

$$z = -z_0\,[3(h^2 x^2 + y^2) - (h^2 x^2 + y^2)^2], \qquad (2)$$

where $h^2$ is close to unity and a function of the incidence and diffraction angles at the grating (Fig. 5)

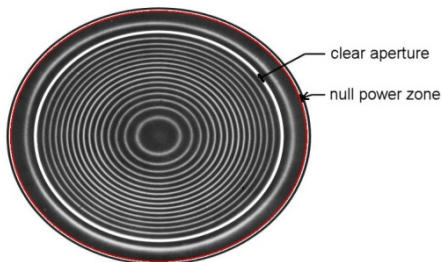 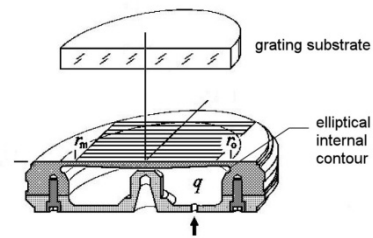

Fig. 5. Freeform grating surface defined by homothetic ellipses.    Fig. 6. Deformable matrix for Fireball-II grating replica.

Similarly to Figure 2, the elasticity design of the active matrix is basically a constant thickness plate *clamped* into an elliptical contour at the null power zone level and bent by a uniform gas pressure $q$ (Fig.6). If $t$ is the thickness of the plate and $D \propto t^{\,3}$ its rigidity, the bilaplacian Poisson's equation, $\nabla^4 z(x,y) - q/D = 0$, provides solutions $t\,(q)$. One selected stainless steel AISI420, $t$=6.7mm, $q$=6.5Atm and an ellipse built-in size 132x140mm.

## 5. Conclusion

Active optics methods provide fascinating imaging quality in astronomy either in high angular resolution or limiting magnitude detection. These methods are key features to generate extremely smooth freeform surfaces.